\documentstyle[12pt]{article}
\begin{document}

\title{ Intercellular Communication Via Intracellular Calcium Oscillations}

\author{M. E. Gracheva and  J. D. Gunton\\
Department of Physics\\
Lehigh University \\
Bethlehem, PA 18015} \maketitle

\begin{abstract}
In this letter we present the results of a simple model for  intercellular
communication via calcium oscillations, motivated in part by a recent experimental
study. The model describes two cells (a "donor" and "sensor") whose intracellular
dynamics involve a calcium-induced, calcium release process. The cells are coupled by
assuming that the the input of the sensor cell is proportional to the output of the
donor cell. As one varies the frequency of calcium oscillations of the donor cell, the
sensor cell passes through a sequence of N:M phase locked regimes and exhibits a
"Devil's staircase" behavior. Such a phase locked response has been seen
experimentally in pulsatile stimulation of single cells. We also study a stochastic
version of the coupled two cell model.  We find that phase locking holds for
realistic choices for the cell volume.
\end{abstract}

\newpage

Oscillatory increases in the intracellular concentration of calcium control a variety
of important, diverse biological functions, including muscle contraction, metabolism
and gene expression \cite{BLB00,BBL98,AG96}. In the latter case, for example, calcium
oscillations lead to the expression of genes that are essential for dendritic
development and neuronal survival. A recent review of the versatility and
universality of calcium signalling has been given by Berridge et al \cite{BLB00}.
Typically cells have a Ca$^{2+}$ rest concentration of about 100 nM, but when
activated rise to concentrations of roughly ten times this. Such increases can be
produced by ligands (agonists) binding to receptors located on the plasma membrane,
through a process involving the second messenger inositol-1,4,5-trisphosphate
(IP$_3$). These can be receptor-specific, as shown in a recent study of the
relationship between the production of IP$_3$ and the calcium response \cite{NYC01}.
An important characteristic of the spike-like Ca$^{2+}$ oscillations is that they are
frequency, rather than amplitude, encoded. That is, an increase in the agonist
concentration increases the frequency of oscillation, but has little effect on its
amplitude. Another significant characteristic is that calcium signals can be
propagated between cells, providing an important means of cell communication.  Such
intercellular communication can take different forms, including diffusion of calcium
or IP$_3$ through gap junctions and paracrine signaling.   Recently deterministic
models have been developed for signalling through gap-junction diffusion, via a
second messenger such as calcium or IP$_{3}$ \cite{D00,H99,H01}. Important stochastic
effects have also been included \cite{GTG01} for gap-junction signalling, as well as
for other aspects of calcium dynamics \cite{BFL00,FTL00,LIX00}. In one type of
paracrine signalling,  a calcium spike in one cell causes the release of a secondary
agonist, such as  ATP, to the extracellular space, followed by stimulation of
purinergic receptors on nearby cells \cite{PRS99,PG99}. Recently a new paracrine
mechanism for intercellular communication has been proposed \cite{HCDB00}, based on
the fact that the calcium liberated as a consequence of intracellular calcium spiking
is often extruded to the extracellular neighborhood of the cell. The recent
experimental study showed if this space is limited such that the local extracellular
calcium fluctuations are sufficiently large, calcium-sensing receptors (CaR)
\cite{CaR00} on the surfaces of adjacent cells can be activated, producing secondary
spikes in these cells. Thus calcium receptors may mediate a new form of intercellular
communication in which cells are informed of the intracellular signaling of their
neighbors via extracellular calcium fluctuations. However, the experimental results
yield only qualitative information about the response of the
sensor cell to the donor cell.  \\
\\
 In this letter we present the results of
a simple model for paracrine intercellular  communication via calcium oscillations,
motivated in part by this recent experimental study. As one still does not understand
in detail the complex biochemistry involved in the CaR coupling, we limit ourselves
to studying a simplified model that might capture the qualitative features of this
new form of signaling. There are two aspects to describing the intercellular
communication: the intracellular dynamics and the coupling between cells.   A number
of theoretical models have been developed to explain  intracellular Ca oscillations
\cite{GDB90,DGB90,CLF95,KOD00}. The basis for most of these is that after an agonist
(hormone) binds to the extracellular side of a receptor bound to the membrane, the
G$_{\alpha}$ subunit at the intracellular side of the receptor-coupled G-protein is
activated. This activated G-protein then stimulates a phospholipase C (PLC) which
helps form a second messenger IP$_3$  and diacylglycerol. IP$_3$ then binds to
specific receptors in the membrane of an internal store of calcium (such as the
endoplasmic reticulum). The binding helps open calcium channels, which leads to a
large flux of calcium ions from the internal store into the cytosol, which then
stimulates the release of additional calcium ions.  Some
details of this complex progress, however, remain unknown.\\
As there are many
different models for the intracellular calcium oscillations, we choose the simplest
to illustrate how the results of communication between cells might differ depending
on the internal cell dynamics. This is the so-called minimal model, involving two
dynamical variables, the cytosolic calcium and an internal store of calcium (such as
the endoplasmic reticulum) respectively \cite{GDB90}, in which an agonist induces
calcium oscillations in a single cell. We couple two such cells, the donor cell and
the sensor cell,  by assuming that the stimulus of the target cell is proportional to
the cytosolic calcium content of the first cell. Since some of the cytosolic
Ca$^{2+}$ produced in the donor cell is extruded into a small space near a CaR
receptor, this seems to be a reasonable assumption. This avoids modeling the
extracellular diffusion of Ca$^{2+}$ as well as the complex receptor dynamics that is
presumably involved in the calcium-sensing receptor mechanism proposed by H\"ofer et
al. \cite{HCDB00}. However, our model is consistent with the spirit of the single
cell minimal model in that it provides a minimal two cell coupling that yields
interesting intercellular communication.  We should also note that  under in vivo
conditions, hormones are not released steadily, but are released in a pulsatile
fashion.  Thus our results for the sensor cell responding to an input signal are in
principle relevant to the physiologically interesting question of how the
intracellular cytosolic calcium responds to a pulsatile application of agonists.
\\

 We  consider a coupled cell version of the minimal model, coupling
 two cells through a term proportional to the calcium output of the first (donor) cell.
This term will serve as an external stimulus for the second (sensor) cell.  The donor
cell dynamics is described by two differential equations for its cytosolic Ca$^{2+}$
concentration, $y_{1}$ and its internal store of Ca$^{2+}$, $y_{2}$:

\begin{eqnarray}
\frac{dy_1}{dt}&=&V_0+ \beta_1 V_1-V_2+V_3+k_f y_2-k y_1\\
\frac{dy_2}{dt}&=&V_2-V_3-k_f y_2
\end{eqnarray}

where  $V_2=\frac{V_{m2} y_1^2}{(k_2^2+y_1^2)}$ and $V_3=\frac{V_{m3} y_1^4
y_2^2}{(k_a^4+y_1^4)(k_r^2+y_2^2)}$.
 This model has been studied extensively \cite{GDB90,DGB90}. It is known
 that for a given set of the parameter values Ca$^{2+}$ oscillations will occur when
 the
 parameter $\beta_1$, which increases with the concentration of the external hormonal stimulus,
 lies in a
 range $\beta_{min}<\beta_1<\beta_{max}$. The minimum and maximum values depend mainly
 on the parameters $V_0$ and $V_1$.\\
The sensor cell is modeled using the same equations for its cytosolic and internal
calcium concentrations $y_{1}^{'}$ and $y_{2}^{'}$ as given in Eqs(1,2).   However,
instead of a term  $\beta_2V'_1$ representing a constant stimulus, we use the term
$\beta_2 y_1V'_1$, which provides the coupling between the cells. This assumes that
the stimulus to the donor cell from the extruded calcium from the donor cell is
proportional to the the latter's cytosolic calcium concentration.
In general, the structural parameters $V_0$, $V_1$, $V_2$, $V_3$, $k_f$, $k$ of the
first cell and $V'_0$, $V'_1$, $V'_2$, $V'_3$, $k'_f$, $k'$ of the second  cell can
be different, but for the sake of simplicity we take them to be the same. We find  in
general that oscillations in the donor cell due to a constant hormonal input produce
oscillations in the sensor cell. This is in qualitative agreement with the
experimental observation \cite{HCDB00}, but the detailed predictions of our model
require further experimental study.\\
We have calculated the N:M rhythms predicted for
this coupled minimal model as a function of $\beta_1$ for fixed $\beta_2$, where N
denotes the number of stimuli arising from the donor cell and M the number of
responses of the sensor cell in a given time interval. For example, the  frequency of
response can be the same as the frequency of the stimulus, i.e. N:M is 1:1. However,
in general the Ca$^{2+}$ response in the sensor cell is blocked when the frequency of
pulses of the donor cell is increased. In Fig. 1 we show a 3:2 response.  This
phenomenon of blocking is also seen in heart patients, where it is known as Wencekbach
periodicity.  As one varies $\beta_1$  the sensor cell passes through a sequence of
N:M phase locked regimes (in response to the oscillatory stimuli from the donor cell)
and exhibits a "Devil's staircase" behavior \cite{RS94}, as shown in Fig. 2. That
is, between any two steps there is a countless number of staircases. This behavior
has been found earlier in a study of a finite difference model of cardiac arrhythmias
\cite{GGS87} as well as in a model of intracellular calcium oscillations \cite{CLF95}
in which the hormonal stimulus was modeled by a sequence of square well pulses.
However, this is the first prediction of such behavior in coupled, nonexcitable
cells.  We find, for example, that with $k=6s^{-1}$, $k'=6s^{-1}$, $\beta_2=0.4$ and
all other parameters as in Table 1 the variation of $\beta_1$ from $0.3$ to
$0.415$, i.e. increase in the concentration of the external stimulus which increases
the frequency output of the first cell, leads to the ratio of the stimulus/response
from 1:1 rhythm ($\beta_1=0.3$), through 5:4 ($\beta_1=0.4$), 4:3 ($\beta_1=0.405$),
3:2 ($\beta_1=0.41$), 5:3 ($\beta_1=0.412$) to the 2:1 rhythm ($\beta_1$=0.415). On
the other hand, various rhythms also can be obtained by fixing, for example,
$\beta_1=0.3$ and varying $\beta_2$ from 0.38 to 0.345 (all other parameters as
described above). In this case we find the following sequence of the
stimulus/response rhythms: 1:1 rhythm ($\beta_2=0.38$), 5:4 ($\beta_2=0.37$), 4:3
($\beta_2=0.365$), 3:2 ($\beta_2=0.36$) and finally a 2:1 rhythm ($\beta_2$=0.345).
Some examples of the Devil's staircase are shown in Fig. 2 and Fig. 3.  This response
of the sensor cell is similar to experimental results of Sch$\ddot{o}$fl et al.
\cite{SBH93} who applied square wave pulses to phenylephrine to liver cells every 30
seconds. They found stimulus/response rhythms such as 2:1, but with less regularity
than shown here \cite{P98}. A subsequent stochastic model based on a deterministic
model of intracellular dynamics due to Chay et al \cite{CLF95} yielded results
qualitatively similar to the experiment \cite{P98}.
\\
Deterministic models such as the one used above neglect potentially important
stochastic effects such as fluctuations in the baseline values of Ca$^{2+}$ and
variations in the amplitudes and widths of the spikes. Since the number densities of
the intracellular signaling molecules are typically low (of the order of $1 - 10^2 \mu
m^{-3}$, stochastic effects could be important. To see whether such effects are
significant here, we have also studied a stochastic version of our model, using a
Monte Carlo method due to Gillespie \cite{DG76}.  We have studied the stochastic model
for different values of the cell volume $\Omega $ (assumed to be the same for both
cells).  For very small $\Omega $ fluctuations destroy the phase locking, while in
the limit of large $\Omega $ one recovers the deterministic limit. Both results are
what one would expect.  For intermediate values of $\Omega $, however, such as
$\Omega = 2000$, which is the approximate volume of hepatocyte cells, we find that
phase locking persists, although with occasional lapses. Some typical results for
this case are shown in Fig. 4. Thus we find a stochastic version of the Devil's
staircase for values of the cell volume that are realistic.  We also found that cells
can switch between frequencies in the stochastic model if we choose $\beta_1$ and
$\beta_2$ such that the deterministic model would give a frequency locking of the cells
on the edge of one of the steps of the Devil's staircase.\\
In conclusion, we have shown that a coupled minimalist model can account for a
variety of calcium oscillations that have been seen experimentally in hepatocytes
stimulated with time-dependent pulses of hormone \cite{SBH93}. This simple model can
also describe intercellular communication between cells via calcium-sensing receptors,
with results that are at least qualitatively consistent with a recent experimental
study \cite{HCDB00} We have found in addition that the deterministic version of the
model yields a Devil's staircase type of phase locking between the donor and sensor
cell. We have also found that this phase locking is present in a stochastic version
of this model, which is a novel feature of the study. The stochastic model is more
realistic than the deterministic model and yields baseline fluctuations and
variations in the amplitude of the spikes in Ca$^{2+}$, as seen in experimental
studies of calcium oscillations. Whether or not paracrine communication in real
biological systems will exhibit a Devil's staircase behavior is an open question,
however, as there are many ways in which one should improve our model to make it more
realistic.  For example, we are currently extending this study to include the plasma
membrane receptor dynamics \cite{ROL99}, and oscillations in $IP_3$ that have been
seen in some recent studies \cite{HKT99,TT01,NYC01}. Our preliminary results from a
study of two cells whose internal dynamics is given by a model of Kummer et al
\cite{KOD00} with coupling through receptor dynamics similar to that of Riccobene et
al \cite{ROL99} show that bursting behavior, in addition to the type of behavior
reported here, is also possible for this form of paracrine communication. The
fundamental problem of paracrine cell communication would seem to be a rich field for
further
experimental and theoretical investigation.\\
\\
{\bf Acknowledgment} This work was supported by NSF grant DMR9813409.

\newpage
{\bf Figure Captions}\\

Fig.1 Calcium oscillations of two connected cells ($\beta_1$=0.3, $\beta_2$=0.36). Frequencies of cells are locked in a sequence of 3:2. Deterministic model.

Fig.2 Devil's staircase, a ratio $N/M$ (where $N$ is the number of spikes of the donor
cell and $M$ is the number of spikes of the sensor cell) as a function of  $\beta_1$
at fixed $\beta_2$=0.3.

Fig.3 Devil's staircase, a ratio $N/M$ as function of  $\beta_1$ at fixed $\beta_2$=0.2.

Fig.4 Calcium oscillations of two connected cells ($\beta_1$=0.17, $\beta_2$=0.3).
Frequencies of cells are locked in a sequence of 1:1 with occasional fluctuations.
Stochastic model with  $\Omega=2000$.

\newpage

\begin{table}[ht]
\begin{center}
\caption{Parameters for the minimal two variable model. \label{table1}} \vspace{5mm}
\begin{tabular}{cc}
 Parameter&Value\\
\hline
$k$&6$s^{-1}$\\
$k_f$&1.0$s^{-1}$\\
$k_2$&1.0$\mu M$\\
$k_a$&0.9$\mu M$\\
$k_r$&2.0$\mu M$\\
$V_0$&1.0$\mu M s^{-1}$\\
$V_1$&7.3$\mu M s^{-1}$\\
$V_{m2}$&65.0$\mu M s^{-1}$\\
$V_{m3}$&500.0$\mu M s^{-1}$\\
$\beta$&$0.1-0.9$\\
\end{tabular}
\end{center}
\end{table}


\begin{thebibliography}{ABCD00}
\bibitem{CaR00} The CaR receptors were initially cloned from the parathyroid gland and
possess the characteristic seven transmembrane segments of the G-protein coupled
receptor family.
\end{thebibliography}
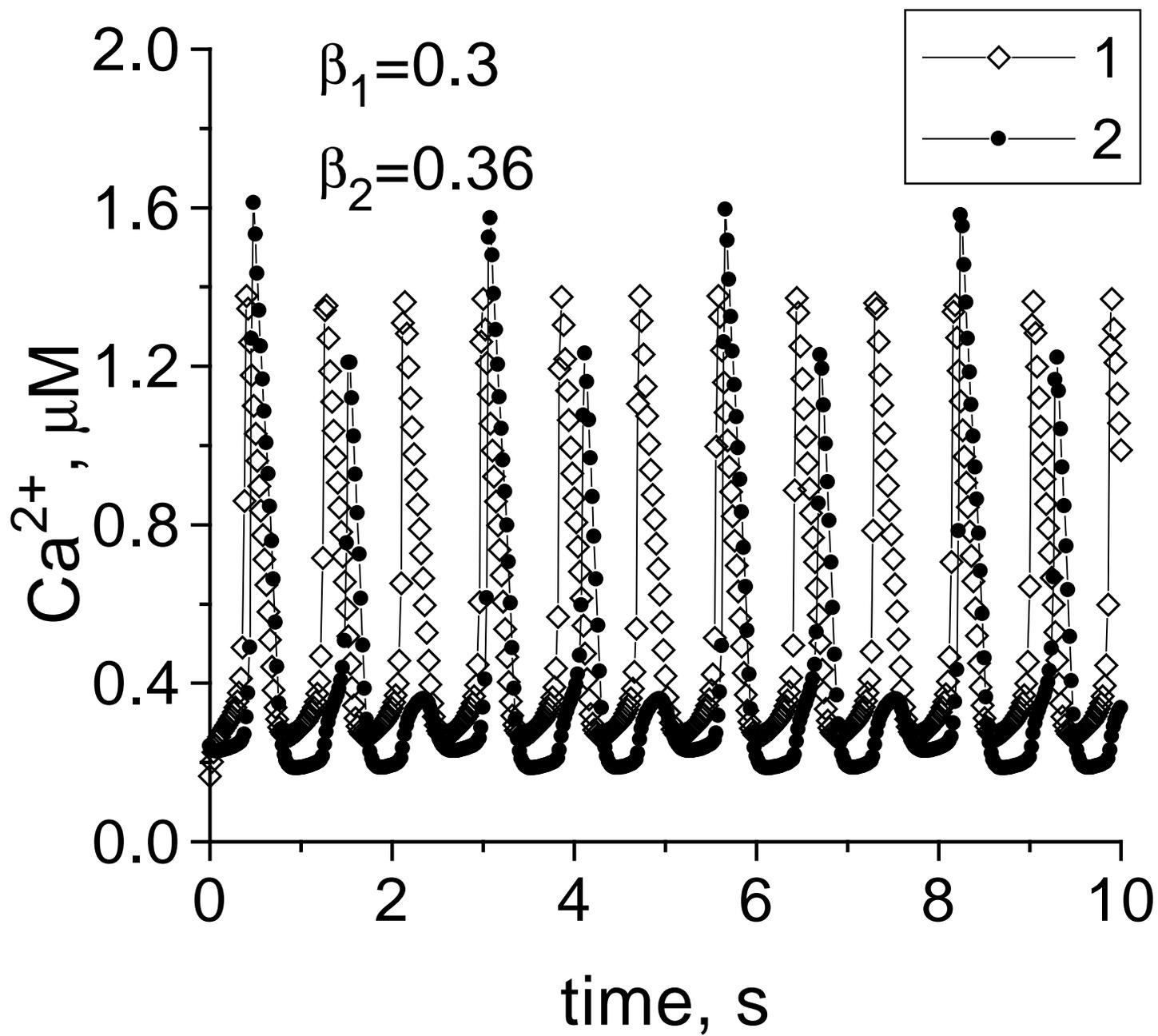
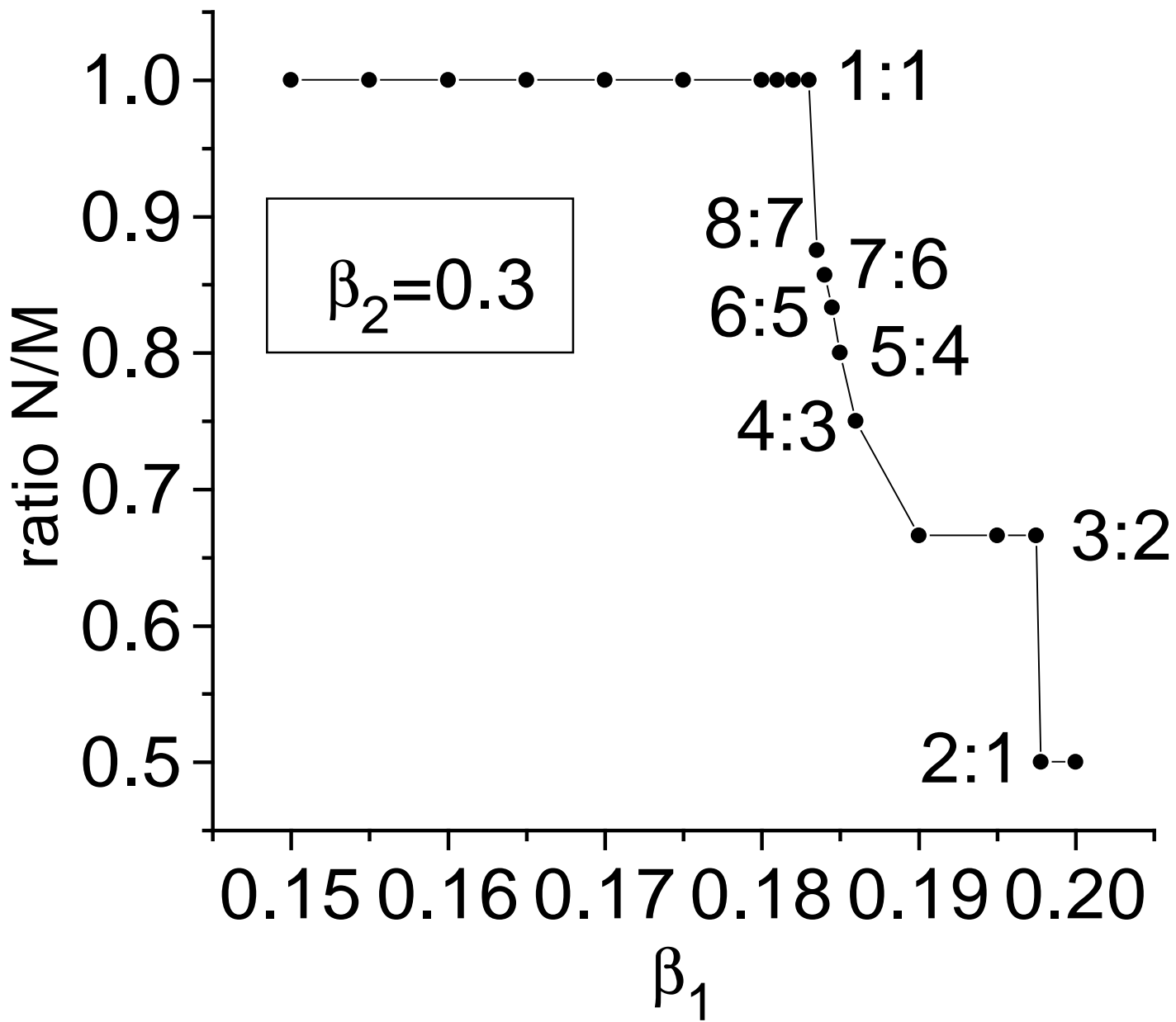
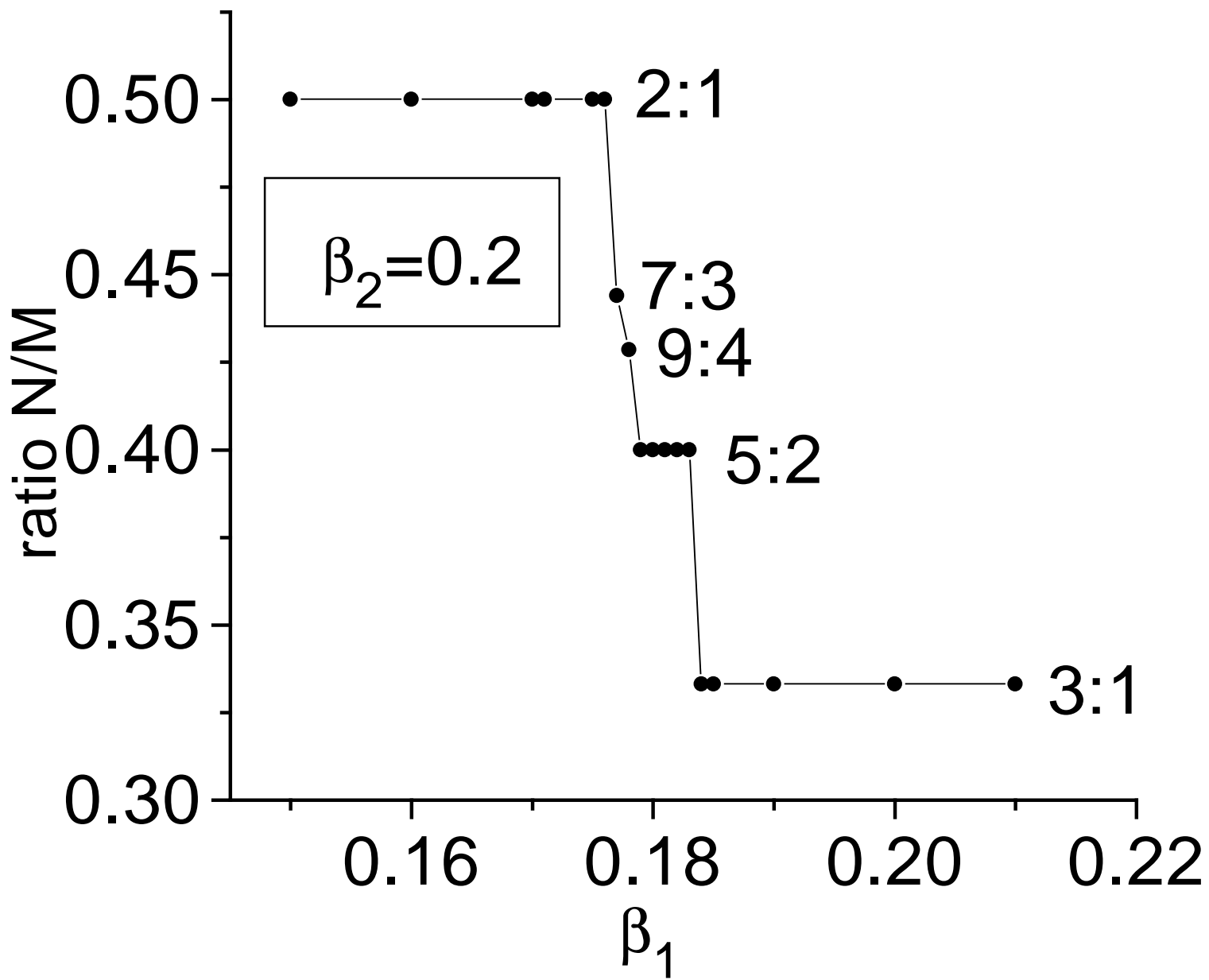
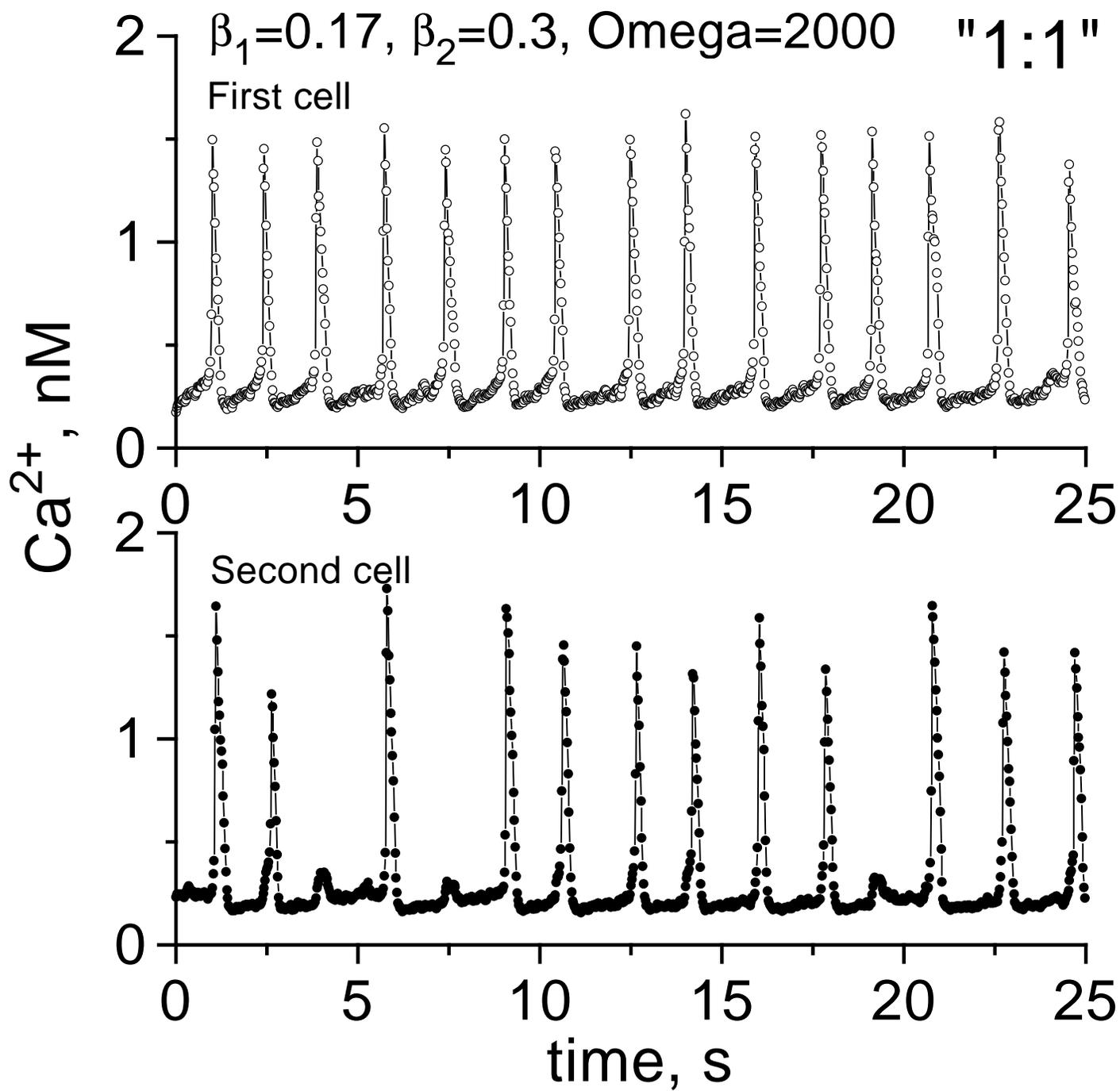
\end{document}